\documentclass[pra,twocolumn,amsmath,amssymb]{revtex4-1} %

\usepackage{epsfig,amsmath}
\usepackage{subfigure}
\usepackage{graphicx}
\usepackage{dcolumn}
\usepackage{stmaryrd}
\usepackage{mathrsfs}
\usepackage{pifont}
\usepackage{amsthm}
\usepackage{amssymb}
\usepackage{bm}
\usepackage{latexsym}
\usepackage[colorlinks=true,linkcolor=blue,citecolor=blue]{hyperref}
\usepackage{color}

\newcommand{\De}{\Delta}

\newcommand{\non}{\nonumber}
\newcommand{\ti}{\tilde}
\newcommand{\la}{\langle}
\newcommand{\ra}{\rangle}

\newcommand{\om}{\omega}
\newcommand{\Om}{\Omega}

\newcommand{\bs}{\bar{1}}

\def\pra#1{{ Phys.\ Rev. A\/} {\bf#1}}
\def\prb#1{{ Phys.\ Rev. B\/} {\bf#1}}

\def\prl#1{{ Phys.\ Rev.\ Lett.} {\bf#1}}

\def\sci#1{{ Science} {\bf#1}}

\def\rmp#1{{ Rev. \ Mod. \ Phys.} {\bf#1}}
\def\nat#1{{ Nature} {\bf#1}}

\def\njp#1{{ New\ J. \ Phys.\/} {\bf#1}}

\begin{document}

\title{Scaling Behavior in the Decoherence of Decoupled Multi-spin System}

\author{Jun Jing$^{1,2}$ \footnote{[Email address]: junjing@jlu.edu.cn} and Xuedong Hu$^{1}$ \footnote{[Email address]: xhu@buffalo.edu}}

\affiliation{$^{1}$Department of Physics, University at Buffalo, SUNY, Buffalo, NY 14260, USA\\ $^{2}$Institute of Atomic and Molecular Physics, Jilin University, Changchun 130012, Jilin, China}

\date{\today}

\begin{abstract}
We study the scaling of decoherence of decoupled electron spin qubits due to hyperfine interaction. For a superposed state consisting of product states from a single Zeeman manifold, both $T_2^*(n)$ and $T_2(n)$ are scale-free with respect to $n$ and the number of basis states, $m$. For a superposed state made up of states from different Zeeman manifolds, both $T_2(n)$ and $T_2^*(n)$ are roughly inversely proportional to $\sqrt{n}$. Our results can be extended to other decoherence mechanisms, including in the presence of dynamical decoupling, which allow meaningful discussions on the scalability of spin-based coherent solid state quantum technology.
\end{abstract}

\pacs{03.65.Yz, 76.30.-v, 71.70.Jp}

\maketitle

{\em Introduction.}---Large-scale quantum information processing (QIP) requires the generation, manipulation, and measurement of fully coherent superposed quantum states involving many qubits \cite{Chuang_book}. One of the key issues for QIP is how well such a many-qubit system can maintain its quantum coherence. This is also an important issue from the perspective of fundamental physics: it remains an intriguing question how a large number of microscopic quantum mechanical systems together behave classically as a macroscopic object. Again, decoherence is central to such quantum-to-classical transitions \cite{Zurek_PhysToday}.

A confined single electron spin in a semiconductor quantum dot (QD) or a shallow donor is highly quantum coherent, and is an ideal candidate as a qubit \cite{Loss_PRA98, Kane_Nature98, Hanson_RMP07, Morton_Nature11, Awschalom_Science13}. At low temperatures, an isolated electron spin has an exceedingly long longitudinal relaxation time \cite{Rashba_PRL03, Golovach_PRL04, Amasha_PRL08, Morello_Nat10}, and a very long pure dephasing time after removing inhomogeneous broadening \cite{Petta_Science05, Bluhm_NP11, Pla_Nat12,Muhonen_Nnano14}. It is now well understood that the main single-spin decoherence channel is through hyperfine coupling to the environmental nuclear spins \cite{Liu_NJP07, Cywinski_PRB09, Bluhm_NP11, Pla_Nat12, Muhonen_Nnano14}, and the effects of hyperfine interaction have also been investigated for coupled two-, three- and even more spin systems \cite{Coish_PRB05, Yang_PRB08, Hung_PRB13, Dial_PRL13, Ladd_PRB12, Medford_PRL13, Mehl_PRB13, Hung_PRB14, Kim_Nat14}. On the other hand, decoherence of a many-spin system is still unexplored.

\begin{figure}[htbp]
\centering
\includegraphics[width=3.2in]{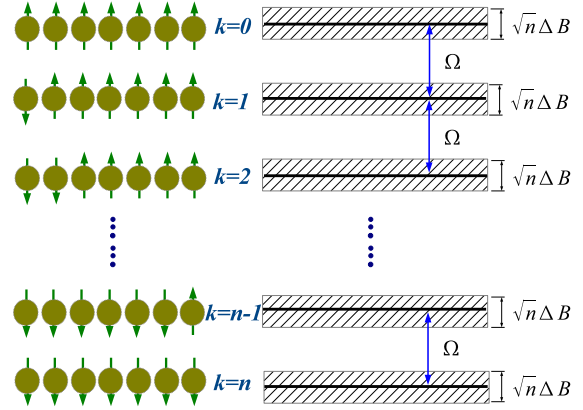}
\caption{The sketch about the energy distribution of manifolds ($0\leq k\leq n$) for $n$ electrons separately confined in $n$ decoupled QDs. Each electron spin is coupled to local nuclear spins by hyperfine interaction that gives rise to local field as large as $\De B$, so that the energy level for each manifold is broadened to a band with width $\sqrt{n}\De B$.}\label{QD}
\end{figure}

In this Letter we study hyperfine-induced decoherence of $n$ $(\gg 1)$ decoupled QD-confined electron spin qubits. Our goals are to clarify how fast a many-qubit superposed state loses its coherence, and how this {\em collective decoherence} scales with the number of qubits involved. In our study, a uniform magnetic field is applied, so that the Zeeman splitting is much larger than the nuclear-spin-induced inhomogeneous broadening (see Fig.~\ref{QD}). Consequently, the dominant single-spin decoherence channel is pure dephasing due to the nuclear spins \cite{Liu_NJP07, Cywinski_PRB09}. We explore how this dephasing mechanism affects a many-spin-qubit state by examining a large number of superposed states in various forms. Our results from this broad-ranged exploration indicate a sublinear scaling behavior for dephasing rates in the short time limit, making the scale-up of a spin-based quantum computer a difficult but not intractable endeavor.

{\em Electron-nuclear spin hyperfine interaction.}---We consider $n$ decoupled electron spins in a finite uniform magnetic field, each confined (in a quantum dot, nominally) and interacting with a local and uncorrelated nuclear-spin bath through hyperfine interaction. The total Hamiltonian for this $n$-qubit system and the nuclear spin reservoirs is
\begin{eqnarray}\non
& & H_{\rm tot}=H_{Z_e}+H_{Z_n}+H_{\rm hf} \\ \label{Hhf}
&=&\sum_{j=1}^n\left(\Om S_j^z+\sum_{\alpha=1}^{N_j} \om_{j \alpha}I_{j \alpha}^z+\sum_{\alpha=1}^{N_j} A_{j \alpha}
\vec{S}_j \cdot \vec{I}_{j \alpha}\right)\,,
\end{eqnarray}
where $\Om$ is the electron Zeeman splitting, $\om_{j \alpha}$ is the nuclear Zeeman splitting of the $\alpha$-th nuclear spin in the $j$-th QD (from here on $j$ will always be used to label the QDs), and $A_{j \alpha}$ is the corresponding hyperfine coupling strength. The number of nuclear spins coupled to each electron spin, $N_j$, is generally large, in the order of $10^5$ to $10^6$ in GaAs QDs, and $\sim 10^3$ in natural Si QDs.

With the electron spins isolated from each other, the total Hamiltonian is a sum of $n$ fully independent single-spin decoherence Hamiltonians. The evolution operator for the $n$-qubit can thus be factored into a simple product of operators for each individual qubits (before and after tracing over the local nuclear reservoirs). We present a brief recap of single-spin decoherence \cite{Liu_NJP07,Cywinski_PRB09} properties in appendix \ref{single}, and focus here on how we approach the multi-spin-qubit decoherence problem based on the results of the single-qubit case. We also note that inhomogeneous broadening and the narrowed-state free induction decay are statistically independent because of independence between longitudinal and transverse Overhauser fields, as presented in appendix \ref{supple}. These two pure dephasing channels follow the same scaling law, i.e., $T_2^*(n)/T_2^*(1) = T_2(n)/T_2(1)$. Thus in the following we will focus on the scaling analysis of $T_2^*(n)$.

{\em Multi-spin decoherence.}---For an $n$-spin system in a finite uniform external magnetic field, the full Hilbert space is divided into $n+1$ Zeeman subspaces, labeled by the expectation value of $S^z=k-n/2$, $k=0,1,2,\cdots,n$.  Each subspace consists of $C_n^k\equiv n!/[k!(n-k)!]$ degenerate states (in the absence of nuclear field), which has $k$ spins in the $|\bs\ra$ state and $n-k$ spins in the $|1\ra$ state. The local random Overhauser field breaks this degeneracy and leads to a broadening of $\sim \sqrt{n}\De B$, as illustrated in Fig.~\ref{QD}. For our decoherence calculations, we use the spin product states $|x_r\rangle = |l^{r}_nl^{r}_{n-1}\cdots l^{r}_1\ra$ as the bases. Here $l_j^r$ refers to the electron spin orientation along the $z$-direction in the $j$th QD for state $|x_r\rangle$, and takes the value of $1$ or $\bs\equiv-1$ for notational simplicity .

For a superposed state $|x\ra$ that contains more than one product state, decoherence emerges due to the non-stationary random phase differences among $m$ product states $|x_r\ra$'s: $|x(t)\ra=\sum_{r=1}^m d_r\exp\{-i\ti{B}_{l^r_nl^r_{n-1}\cdots l^r_1}^zt\}|x_r\ra$ with $\sum_{r=1}^m|d_r|^2=1$ (from now on we use $m$ to represent the number of product states contained in $|x\ra$, and the notation for the Overhauser field is defined in the appendix \ref{notation}). As a collective decoherence measure of $|x\ra$ caused by the inhomogeneous broadening [i.e., in the following calculations we use only the longitudinal Overhauser field $B$ instead of the total Overhauser field $\ti{B}$], we use fidelity defined as $\mathcal{F}(t)=\sqrt{M[\la x|x(t)\ra\la x(t)|x\ra]}$ [see Eqs.~(\ref{dFt}) and (\ref{Ft})]. For $|x\ra$, we find
\begin{equation}\label{gFt}
\mathcal{F}(t)=\sqrt{\sum_{r=1}^m|d_r|^4+2\sum_{i<k}|d_i d_k|^2 M[\cos(\theta_{ik}t)]}\,,
\end{equation}
where the phase difference is $\theta_{ik}\equiv B_{l^i_nl^i_{n-1}\cdots l^i_1}^z-B_{l^k_nl^k_{n-1}\cdots l^k_1}^z$. Specifically, $M[ \cos (\theta_{ik}t) ]$ is solely determined by the number $j_{ik}$ of spins that are opposite in orientation between basis states $|x_i\ra$ and $|x_k\ra$. For example, if $|x\ra=(|11\bs\ra+|1\bs1\ra+|\bs11\ra)/\sqrt{3}$, then $j_{ik}=2$, so that Eq.~(\ref{gFt}) takes on the form $\mathcal{F}(t) = \sqrt{3 + 2M[\cos\theta_{12}t + \cos\theta_{13}t + \cos\theta_{23}t]}/3$, where $\theta_{ik}$ happens to be $2(B_i^z-B_k^z)$. After a semiclassical evaluation of the Overhauser field noise, and using the result of $T_2^*(1)$ in Eq.~(\ref{T2star}), we find $M[\cos\theta_{ik}t] = M[e^{2i(B_i^z-B_k^z)t}] = e^{-8[t/T^*_2(1)]^2}$, so that $\mathcal{F}(t)\approx\exp\{-8/3[t/T^*_2(1)]^2\}$ in the short time limit. Thus in this example, $T_2^*(3)=T_2^*(1)\sqrt{3/8}$, with $\mathcal{F}(t)\equiv\exp\{-[t/T_2^*(n)]^2\}$.

{\em Examples of multi-spin decoherence.}---With our understanding of single-spin decoherence, and with a measure (fidelity) of the collective decoherence for a multi-spin state $|x\ra$, we are now in position to clarify the scaling of the inhomogeneous broadening time $T_2^*(n)$ in various subspaces of the $n$-spin system. Below we describe the results from several representative classes of $|x\ra$.

{\em Case A: single product state.}---The simplest multi-spin state is a single product state ($m=1$). The random Overhauser field acting on a product state creates a random but global phase (relative to when the nuclear reservoir is absent). This global phase does not lead to any decoherence, as there is no coherence (phase) information stored in a product state to begin with.

{\em Case B: two product states, with $m=2$ and $k\geq1$.}---The simplest multi-spin state that can undergo pure dephasing consists of two product states.  Here we choose a particular class of $|x\ra_B = d_1|b\ra+d_2|k\ra$, with one product state being fully polarized $|b\ra=|1\ra^{\otimes n}$, while the other being from the $k$-th subspace with $k$ spins prepared in $|\bs\ra$. The fidelity of such a state is given by
\begin{equation}
\mathcal{F}(t)\approx\exp\left\{-4|d_1 d_2|^2k \left[\frac{t}{T_2^*(1)}\right]^2\right\} \,,
\end{equation}
so that
\begin{equation}\label{Ca}
\frac{T_2^*(n)}{T_2^*(1)}=\frac{1}{2|d_1 d_2|\sqrt{k}}\,.
\end{equation}
In this case, dephasing time is inversely proportional to the square root of the number of spins prepared as $|\bs\ra$ in $|k\ra$. A special example here is the GHZ state, $|x\ra_{\rm GHZ} = (|1\ra^{\otimes n} + |\bs\ra^{\otimes n})/\sqrt{2}$, for which the two product states have completely opposite spins. The decoherence rate is simply $T_2^*(n)/T_2^*(1) = 1/\sqrt{n}$, where $n$ is the number of spin qubits involved. Indeed, the worst case of scenario for a two-product-state $|x\ra$ is when the pair of product states have completely opposite spins, $T_2^*(n)/T_2^*(1) = 1/\sqrt{n}$.

{\em  Case C: $n\geq m\geq2, k=1$}---We now consider an $|x\ra$ that is a general superposition of $m$ product states from the first manifold with one spin in $|\bs\ra$. In other words, $|x\ra_C=d_1|111\cdots\bs\ra+d_2|11\cdots\bs1\ra+\cdots+d_n|\bs11\cdots\ra$, where $\sum_{j=1}^n|d_j|^2=1$. This is a state that is slightly more general than the $W$-state, with a random weight and phase for each basis state. The fidelity of $|x(t)\ra_C$ is
\begin{equation}
\mathcal{F}(t)\approx\exp\left\{-8\sum_{j_1<j_2}|d_{j_1} d_{j_2}|^2 \left[\frac{t}{T_2^*(1)}\right]^2\right\} \,,
\end{equation}
which implies (by the Cauchy$-$Schwarz inequality)
\begin{equation}\label{sn1}
\frac{1}{2}\sqrt{\frac{n}{n-1}} \leq \frac{T_2^*(n)}{T_2^*(1)}=\sqrt\frac{1}{4-4\sum_{j=1}^n|d_j|^4}\leq\infty\,.
\end{equation}
Here the upper bound ($\infty$ means no decoherence) is approached when a particular $|d_{j_1}|=1$ while all other $|d_{j_2 \neq j_1}|=0$, so that we go back to a single product state. The lower bound corresponds to the equally-populated superposed states with $|d_j|^2=1/n$, i.e., an almost normal $W$-state (a standard $W$ state would have all $d_j$ having the same phase, too). When $n \rightarrow\infty$, $T_2^*(n)/T_2^*(1) \geq 1/2$. The whole system acts like a giant spin$-1/2$ system that is spread out over $n$ physical spins. Notice that the lower bound of decoherence time is scale-free. The scaling of decoherence for large $n$ is insensitive to either the population distribution on each basis state or the total number of physical spins.

{\em Case D: $m=C_n^k, k\geq2$.}---We now extend $|x\ra$ to a more generalized $W$-state that is uniformly distributed over all the product bases in the $k$-th Zeeman manifold, with $k \geq 2$. Consider a particular example, $|x\ra_D = \sum_{r=1}^md_r|x_r\ra$, where $|d_r|^2=1/C_n^k$ and each $|x_r\ra$ has $k$ spins in $|\bs\ra$ and $n-k$ spins in $|1\ra$. The overall decoherence is determined by the phase differences between every pair of states from the $C_n^k$ basis states. Since $C_n^k=C_n^{n-k}$, we limit our discussion below to $k\leq n/2$ without loss of generality. The phase difference $\theta_{r_1 r_2}$ between a particular pair of product states $|x_{r_1}\ra$ and $|x_{r_2}\ra$ can involve Overhauser fields in $2j$ QDs, where $j\leq k$. In the extreme case of $2j=n$, the pair of states has completely opposite spins. For each possible $j$, there are $C_{n-k}^j C_k^{k-j}$ pairs of states with the same phase difference as well as the same value of ensemble average $M[e^{i\theta^j_{r_1 r_2}t}] = \exp \{ -2j [2t/T_2^*(1)]^2 \}$. Therefore the fidelity for this generalized $W$-state is
\begin{eqnarray}\non
\mathcal{F}(t)& = & \frac{1}{C_n^k}\sqrt{C_n^k + 2 \sum_{r_1<r_2} M[\cos(\theta^j_{i_1 i_2}t)]}  \\ \non
& = & \frac{1}{C_n^k} \sqrt{C_n^k +2\frac{C_n^k}{2}\sum_{j=1}^k C_{n-k}^j C_k^{k-j} M[e^{i\theta^j_{r_1 r_2}t}]}  \\
& \approx & \exp\left\{-\frac{4k(n-k)}{n} \left[\frac{t}{T_2^*(1)}\right]^2\right\}\,.
\end{eqnarray}
Thus
\begin{equation}\label{Cc}
\frac{T^*_2(n)}{T^*_2(1)}=\frac{1}{2}\sqrt{\frac{n}{k(n-k)}}\,.
\end{equation}

\begin{figure}[htbp]
\centering
\includegraphics[width=3.2in]{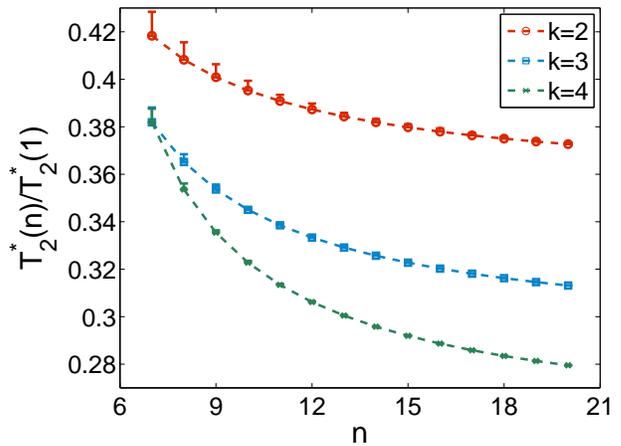}
\caption{The result of $T^*_2(n)/T^*_2(1)$ by randomly generated $|x\ra_D'$ states in the $k$-th manifold comparing to the analysis expression by the generalized $W$-state in {\em Case D}.  As compared to the equally-populated $|x\ra_D$ state, a $|x\ra_D'$ has random populations in each product basis state.}\label{errD}
\end{figure}

In this case, we find that (a) when $n \rightarrow \infty$, $T_2^*(n) \approx 1/(2\sqrt{k})$, which is scale-free with respect to the number of spins $n$ as well as the number of product states $m$ in $|x\ra_D$ (it is a similar feature as in {\em Case C}, where $k = 1$); (b) overall decoherence is completely suppressed when $k = 0$ or $k = n$, i.e. $T_2^*(n=k)=\infty$. These two Zeeman manifolds contain one state each, so that {\em Case D} is reduced to {\em Case A}; (c) the strongest decoherence occurs when $k \approx n/2$, where $T_2^*(n) \approx T_2^*(1)/\sqrt{n}$; (d) the generalized $W$-state $|x\ra_D$ here is a reliable lower bound for the decoherence scaling rate of a more general state $|x\ra_D'=\sum_{r=1}^md_r|x_r\ra$ in the $k$-th manifold. In Fig.~\ref{errD}, the dashed lines represent the analytical result given in Eq.~(\ref{Cc}) with $7\leq n\leq20$ and $2\leq k\leq4$, and the data for error bars are obtained by the maximal errors numerically evaluated on $100$ $|x\ra_D'$ states randomly generated in the $k$-th manifold, with completely random coefficient for each basis state. Figure~\ref{errD} clearly shows that deviations from the result of $|x\ra_D$-state quickly decrease with increasing $n$ and $k$. Thus the equal-weight $|x\ra_D$ state is a very good representative of both {\em Cases C} and {\em D}.

{\em Case E: $m=n$}.---Next we further generalize $|x\ra$ to be a superposition over product states picked from more than one Zeeman manifold. Out of the infinite number of possible combinations, we pick one class of such states, with one product state from each Zeeman manifold, so that $m=n$ and $|x\ra_E = \sum_{r=1}^nd_r|x_r\ra$, with $|x_r\ra$ picked from the $r$-th manifold. To obtain analytical results, we first assume equal weight for all the states involved: $|d_r|^2 = 1/n$. To further limit the choice of states, we assume there is $j$ spin polarization difference between $|x_{r}\ra$ and $|x_{r+j}\ra$, $r=1,2,\cdots,n-j$. For example, for a $3$-spin system, $|x\ra$ can be chosen as $(|\bs11\ra+|\bs\bs1\ra+|\bs\bs\bs\ra)/\sqrt{3}$. For an arbitrary $n$, the fidelity is
\begin{eqnarray}\non
\mathcal{F}(t)&=&\frac{1}{n}\sqrt{n+2\sum_{j=1}^{n-1}(n-j)
\exp\left\{-j\left[\frac{2t}{T_2^*(1)}\right]^2\right\}}\\ &\approx& \exp\left\{-\frac{2(n^2-1)}{3n}\left[\frac{t}{T_2^*(1)}\right]^2\right\}\,.
\end{eqnarray}
Therefore
\begin{equation}\label{caseE}
\frac{T_2^*(n)}{T_2^*(1)}=
\sqrt{\frac{3n}{2(n^2-1)}}\approx\sqrt{\frac{3}{2n}}\,.
\end{equation}

\begin{figure}[htbp]
\centering
\includegraphics[width=3.2in]{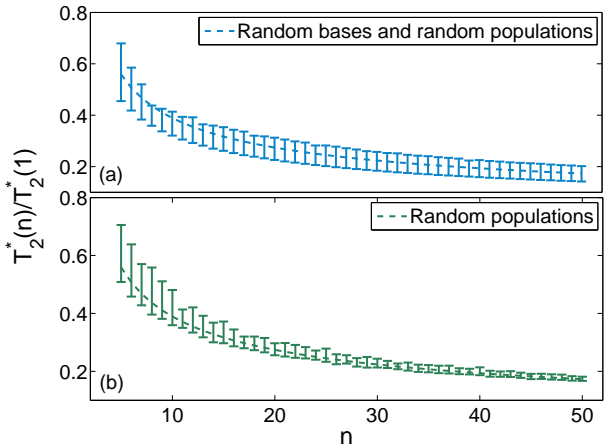}
\caption{The results of $T^*_2(n)/T^*_2(1)$ by randomly generated $|x\ra_E'$ states over all the Zeeman manifolds, as compared with Eq.~(\ref{caseE}) in {\em Case E}. In frame (a), we use $100$ states with random product bases and random populations; in frame (b), we use $100$ states with the fixed bases of $|x\ra_E$ but random populations in each product basis state.}\label{errE}
\end{figure}
As in {\em Case D}, we generalize $|x\ra_E$ to $|x\ra_E'$ by randomizing the weight $|d_r|^2$'s, $1\leq r\leq m$, and the selection of the product basis states within each manifold. In Fig.~\ref{errE} we plot our numerical results as compared with the analytical expression from Eq.~(\ref{caseE}). While the error bars in Fig.~\ref{errE} for random $|x_r\ra_E'$ are larger than those in Fig.~\ref{errD}, the analytical result is still a good indicator of the average $T_2^* (n)$. The size of the error bars also decreases with increasing $n$. Thus the $n$-spin dephasing time scales as $1/\sqrt{n}$ for a large $n$.

{\em Case F: $m=2^n-1$.}---Suppose $|x\ra$ is a combination of {\em Cases D} and {\em E}: $|x\ra_F=\sum_{k=0}^n \sqrt{C_n^k/2^n} |W_k\ra$, where $|W_k\ra$ is a normalized $W$-state in the $k$-th manifold. In fact, $|x\ra_F$ is just the fully superposed state $[(|1\ra+|\bs\ra)/\sqrt{2}]^{\otimes n}$. For the overall decoherence, $C_{2^n}^2$ pairs of phase differences have to be taken into account. There are $2^nC_n^j$ elements involved with $j$ spins, $1\leq j\leq n$, in the set of $\{\theta^j_{i_1i_2}\}, i_1\neq i_2$. The fidelity is
\begin{equation}
\mathcal{F}(t)\approx \exp\left\{-n\left[\frac{t}{T_2^*(1)}\right]^2\right\}.
\end{equation}
Consequently, we have
\begin{equation}\label{T2F}
\frac{T_2^*(n)}{T_2^*(1)}\approx\frac{1}{\sqrt{n}}.
\end{equation}
Based on this equation and the numerical simulation over $|x\ra_F'$ with randomized coefficients as seen in Fig. \ref{errF}, the dephasing time for $|x\ra$ residing in the whole Hilbert space adhere to the sublinear power-law $n^{-1/2}$, the same as in {\em Case E}.

{\em Conclusions and Discussions.}---We have explored the scaling behavior of the decoherence time of $n$ decoupled electron spin qubits by investigating the fidelity of $6$ classes of representative superposed states $|x\ra$. Each electron spin is individually coupled with its own nuclear spin bath through hyperfine interaction, and we do not consider electron-electron interactions in this study.

\begin{table}[htbp]\centering
\begin{tabular}{|c|c|} \hline
$|x\ra$  & $T_2^*(n)/T_2^*(1)$ {\rm or} $T_2(n)/T_2(1)$ \\ \hline
Stable: {\em A} & no decoherence \\ \hline
Two product states: {\em B} & $k^{-\frac{1}{2}}$  \\ \hline
$k$-th subspace: {\em C} and {\em D} & $k^{-\frac{1}{2}}$ \\ \hline
Crossing subspaces: {\em E} and {\em F} & $n^{-\frac{1}{2}}$  \\ \hline
\end{tabular}
\caption{Various scaling behaviors of decoherence time of the $n$  decoupled-electron-spin system under the influence of hyperfine coupling with local nuclear baths.}
\label{scale}
\end{table}

Our results are summarized in Table \ref{scale}, where $k$ is the number of spins in $|\bs\ra$ in a product state that makes up of $|x\ra$.  Typically, both inhomogeneous broadening dephasing rate $1/T_2^*(n)$ and pure dephasing rate $1/T_2(n)$ are sublinear power-law functions of spin number $n$. If $|x\ra$ is constrained in a single subspace with a fixed $k$, $T_2^*(n)$ and $T_2(n)$ become scale-free with respect to $n$ and $m$ (the number of basis states involved).

The scaling behaviors revealed in our case studies can be qualitatively understood based on counting the number of different spin orientations in any pair of product states. Considering any $m$ product states making up a $|x\ra$, a large fraction of pairs have $O(n)$ electron spins oriented in the opposite direction. If we average over all possible states assuming $|d_r|^2 \approx 1/m$, then the state fidelity given in Eq.~(\ref{gFt}) could be estimated as
\begin{eqnarray*}
\mathcal{F}(t)&\approx&\sqrt{m\frac{1}{m^2}+2\frac{1}{m^2}\frac{m(m-1)}{2} e^{-O(n)[t/T_2^*(1)]^2}} \\ &\approx& \exp\left\{-\frac{m-1}{m}O(n)\left[\frac{t}{T_2^*(1)}\right]^2\right\}\, .
\end{eqnarray*}
The decoherence rates are insensitive to $m$ because of normalization and our equal-population assumption. Furthermore, in the $k$-th manifold, the scaling law is $1/\sqrt{k}$ because an arbitrary pair of states is different in $O(k)$ spins.

Our study here could be straightforwardly extended to other decoherence mechanisms. If the single-spin decoherence function is given by $W(t) = \exp\{-[t/T_2(1)]^\nu\}$, the index of every power-law ($-1/2$) in Table \ref{scale} should be revised to $-1/\nu$. For example, spin relaxation induced by electron-phonon interaction produces a linear exponential decay characterized by $T_1$, with $1/T_2=1/(2T_1)$. In this case the scaling power-laws for the $n$-spin system will be modified to be proportional to $k^{-1}$ or $n^{-1}$ based on the selection of $|x\ra$. For decoherence due to Gaussian noise under dynamical decoupling~\cite{Lukasz_PRB08}, the decay functions have $\nu=4$ for spin echo (SE) and $\nu=6$ for two-pulse Carr-Purcell-Meiboom-Gill sequence, so that the scaling factors for decoherence times of the $n$-spin system become $n^{-1/4}$ and $n^{-1/6}$, respectively.

Our results are important to the scale-up considerations for spin-based quantum computers or more general qubits that are under the influence of local reservoirs. The sublinear scaling shows that a large superposed state does not lose its fidelity overly quickly as conventional wisdom may dictate. The scale-free states also help us identify what Hilbert subspaces are more favorable in coherence preservation.

We acknowledge financial support by US ARO (W911NF0910393) and NSF PIF (PHY-1104672). J. J. also thanks support by NSFC grant No. 11175110.

\appendix

\section{Single-spin Decoherence}\label{single}

For a single electron spin coupled to the surrounding nuclear spins in a finite magnetic field, the nuclear reservoir causes pure dephasing via the effective Hamiltonian \cite{Liu_NJP07, Cywinski_PRB09}
\begin{eqnarray}\non
H_{\rm hf}&=&2S^z(H_A+V) \\ \label{SW}
&=&2S^z\left(\sum_{\alpha=1}^N\frac{A_\alpha}{2}I_\alpha^z+\sum_{\alpha \neq \alpha'}\frac{A_\alpha A_{\alpha'}}{4\Om}I_\alpha^+I_{\alpha'}^-\right)\,.
\end{eqnarray}
where $N$ is the number of nuclear spins, $\Om$ is the electron Zeeman splitting, and $A_\alpha$ is the hyperfine coupling strength. The sums over $\alpha$ and $\alpha'$ here are over all the nuclear spins in the single quantum dot (QD). The dephasing dynamics has two contributions: $H_A$ is the longitudinal Overhauser field, while $V$ is the second-order contribution from the transverse Overhauser field. In a finite field, normally the former dominates, generating a random effective magnetic field of $\De B\sim 1 \ {\rm to} \ 5$ mT~\cite{Petta_Science05} on a quantum-dot-confined electron spin in GaAs. This random field leads to a stochastic phase and accounts for the inhomogeneous broadening effect characterized by a free induction decay at the time scale of $T_2^*(n)$, with $n$ being the number of single-electron QDs. For a single dot $n=1$, the inhomogeneous broadening decoherence function is:
\begin{equation}\label{T2star}
M\left[e^{-i\sum_{\alpha=1}^NA_\alpha I_\alpha^zt}\right]=W^A(t)\equiv \exp\left[ -\frac{t^2}{T^*_2(1)^2} \right]\,.
\end{equation}
Here $M[\cdot]$ is an ensemble average over the longitudinal Overhauser field in the QD, and $T_2^*(1) \propto \sqrt{N} / A$ with $A = \sum_\alpha A_\alpha$. In a single gated QD in GaAs, $T^*_2(1)$ is in the order of $10$ ns.

If the effect of $H_A$ is suppressed, such as through nuclear spin pumping and polarization \cite{Bluhm_NP11}, $V$, which is second order in the transverse Overhauser field, leads to the so-called narrowed-state free induction decay (FID), by which the off-diagonal elements of the spin density matrix decay at the time scale of $T_2^{n{\rm FID}}$. In the main text we simplify the notation for $T_2^{n{\rm FID}}$ to $T_2(n)$, in the same way as $T_2^*(n)$. The narrowed-state decoherence function for a single dot is:
\begin{equation}\label{WV}
\left|W^V(t)\right| \approx \exp\left[-\frac{t^2}{T_2(1)^2}\right]\,,
\end{equation}
where $T_2(1) \sim N \Om/A^2$ \cite{Cywinski_PRB09}, and is in the order of $\mu$s in a gated GaAs QD.

\section{Notations on the Overhauser fields}\label{notation}

A convenient way to understand the effect of hyperfine interaction on the $n$-decoupled-qubit system [see Eq.~(\ref{Hhf})] is to introduce the semiclassical Overhauser field: $B^p_{l_nl_{n-1}\cdots l_1}=\sum_{j=1}^n l_j B_j^p$, where $p=z, +, -$ refers to the longitudinal and transverse directions, $l_j$ takes the value of $1$ or $\bs$, and $B_j^p \equiv \sum_\alpha A_{j \alpha} I_{j \alpha}^p$ is the Overhauser field in the $j$th QD. In a finite field and up to second order, the hyperfine Hamiltonian could be diagonalized on the product state basis into
\begin{equation}\label{tiH}
\ti{H}_{\rm hf}=\frac{1}{2}{\rm diag}\left[\ti{B}_{11\cdots1}^z, \cdots, \ti{B}_{\bs\bs\cdots\bs}^z\right] \,,
\end{equation}
where
\begin{equation}\label{correction}
\ti{B}_{l_nl_{n-1}\cdots l_1}^z=B_{l_nl_{n-1}\cdots l_1}^z+\frac{1}{2\Om}
\sum_{j=1}^nl_jB_j^{p_j}B_j^{-p_j} \,.
\end{equation}
Here $p_j=+$ ($-$) if $l_j=1$ ($\bs$). The two terms in Eq.~(\ref{correction}) are responsible for the inhomogeneous broadening and narrowed-state FID, respectively. Accurate to the first order in $A_{j\alpha}$, $B_j^-B_j^+\approx B_j^+B_j^-$ since $[B_j^+, B_j^-]=2\sum_\alpha A_{j\alpha}^2 I_{j\alpha}^z$ is second order in the hyperfine coupling strength and is small.  For simplicity we take $p_j=+$ in the following derivation. Generally, the correction term for the $j$th dot in Eq.~(\ref{correction}) $\approx l_jB_j^+B_j^-/(2\Om)$. For example, a completely polarized state $|b\ra\equiv|1\ra^{\otimes n}$ experiences a longitudinal Overhauser field $\ti{B}_b^z = B_b^z +\sum_{j=1}^nB_j^+B_j^-/(2\Om)$.

Now that the hyperfine Hamiltonian takes on a diagonal form, it can only lead to dephasing between different product states due to $\tilde{B}$, similar to the single-spin case we discussed above. The dephasing of a product state $|x_r\rangle$ relative to $|x_{r'}\ra$ is due to the difference in the random Overhauser field $\tilde{B}$ for these states.

\section{Statistical independence of inhomogeneous broadening and narrowed-state free induction decay}\label{supple}

To analyze the relationship between inhomogeneous broadening and narrowed-state free induction decay in an $n$-decoupled-qubit system, we consider an arbitrary pure state in a subspace spanned by $m$ spin product states $|x\ra=\sum_{r=1}^md_r|x_r\ra$, where $|x_r\rangle = |l^{r}_nl^{r}_{n-1}\cdots l^{r}_1\ra$. Here $l_j^r$ refers to the electron spin orientation along the $z$-direction in the $j$th QD for state $|x_r\rangle$, and takes the value of $1$ or $\bs\equiv-1$ for notational simplicity. This selection is general enough to cover all the cases discussed in the main text. Helped by the Overhauser fields defined above, and under the diagonalized hyperfine interaction Hamiltonian in Eq.~(\ref{tiH}), an initial state $|x\ra$ evolves into
\begin{equation}
|x(t)\ra=\sum_{r=1}^md_re^{-i\ti{B}^z_{x_r}t}|x_r\ra,
\end{equation}
where $\ti{B}^z_{x_r}\equiv\ti{B}_{l^r_nl^r_{n-1}\cdots l^r_1}^z$.  Collective decoherence emerges due to the non-stationary random phase differences among the $m$ product states $|x_r\ra$'s. The fidelity between $|x(0)\ra$ and $|x(t)\ra$ can be expressed as
\begin{eqnarray}\label{dFt}
\mathcal{F}(t)&\equiv&\sqrt{M[\la x(0)|x(t)\ra\la x(t)|x(0)\ra]} \\ \non &=&\sqrt{M\left[\left(\sum_{r=1}^m|d_r|^2e^{-i\ti{B}^z_{x_r}t}\right)
\left(\sum_{r=1}^m|d_r|^2e^{i\ti{B}^z_{x_r}t}\right)\right]} \\ \label{Ft} &=&\sqrt{M\left[\sum_{r=1}^m|d_r|^4
+2\sum_{k<r}|d_k|^2|d_r|^2\cos\theta_{kr}t\right]},
\end{eqnarray}
where the phase differences $\theta_{kr}\equiv\ti{B}^z_{x_k}-\ti{B}^z_{x_r}$.
According to Eq.~(\ref{correction}), each $\theta_{kr}$ could be decomposed into two terms, $\theta_{kr}^{ib}$ and $\theta_{kr}^{ns}$, that are responsible for the inhomogeneous broadening and narrow-state free induction decay, respectively:
\begin{eqnarray*}
\theta_{kr}&=&\theta_{kr}^{ib}+\theta_{kr}^{ns}, \\
\theta_{kr}^{ib}&=&B_{l_n^kl_{n-1}^k\cdots l_1^k}^z
-B_{l_n^rl_{n-1}^r\cdots l_1^r}^z=\sum_{j=1}^n(l_j^k-l_j^r)B_j^z, \\
\theta_{kr}^{ns}&\approx&\frac{1}{2\Om}\sum_{j=1}^n(l_j^k-l_j^r)B_j^+B_j^-.
\end{eqnarray*}
The ensemble average $M[\cos\theta_{kr}t]=M[e^{i\theta_{kr}t}]$ could be estimated using the decoherence times of a single qubit system  $T_2^*(1)$ (inhomogeneous broadening time scale) and $T_2(1)$ (the narrowed-state FID time scale) \cite{Cywinski_PRB09},
\begin{eqnarray}\non
&& M[\cos\theta_{kr}t]=M\left[e^{i\theta_{kr}^{ib}t}
e^{i\theta_{kr}^{ns}t}\right] \\ \non
&=&M\left[e^{i\theta_{kr}^{ib}t}\right]M\left[e^{i\theta_{kr}^{ns}t}\right]
\\ \label{br} &=&\exp\left\{-\sum_{j=1}^n(l_j^k-l_j^r)^2
\left[\frac{t^2}{T_2^*(1)^2}+\frac{t^2}{T_2(1)^2}\right]\right\}.
\end{eqnarray}
This result is obtained by the quantum noise theory \cite{Gardiner_Book}, which is valid at least in the short time limit. Physically it is based on the assumption that longitudinal and transverse Overhauser fields are independent from each other, so that the averages above can be factored. The two decoherence mechanisms are thus mutually independent. Using the short notations $B_{kr}\equiv\sum_j(l_j^k-l_j^r)^2$, $D^{ib}(t)\equiv[t/T_2^*(1)]^2$, and $D^{ns}(t)\equiv[t/T_2(1)]^2$, Eq.~(\ref{Ft}) can be rewritten as
\begin{widetext}
\begin{eqnarray}\non
\mathcal{F}(t)&\approx&\sqrt{\sum_{r=1}^m|d_r|^4
+2\sum_{k<r}|d_k|^2|d_r|^2\left[1-B_{kr}D^{ib}(t)-B_{kr}D^{ns}(t)\right]} \\ \non &=&\sqrt{1-\left(
2\sum_{k<r}|d_k|^2|d_r|^2B_{kl}\right)[D^{ib}(t)+D^{ns}(t)]}
\equiv\sqrt{1-\mathcal{B}[D^{ib}(t)+D^{ns}(t)]} \\ \label{AB}
&\approx&\exp\left[-\frac{\mathcal{B}}{2}D^{ib}(t)\right]
\exp\left[-\frac{\mathcal{B}}{2}D^{ns}(t)\right],
\end{eqnarray}
\end{widetext}
where $\mathcal{B}\equiv2\sum_{k<r}|d_k|^2|d_r|^2B_{kr}$. In short, Eqs.~(\ref{br}) and (\ref{AB}) prove that inhomogeneous broadening and narrowed-state FID are independent decoherence channels, and have the same scaling behavior. The overall decoherence function is just a simple product of the decay functions for inhomogeneous broadening FID and narrowed-state FID. We can thus focus on just inhomogeneous broadening in our discussion of decoherence scaling for $n$ spin qubits and in main text,  we omit the superscript of the phase difference $\theta$ for notation simplicity.  

\section{Numerical evaluation of $n$-spin decoherence}\label{NECaseF}

Equation~(\ref{AB}) gives a general description of decoherence function within the Overhauser field approach. Notice that the fidelity $\mathcal{F}(t)$ of a pure state $|x\ra$ is solely dependent on the function $\mathcal{B}$, which is only a function of populations in product basis $|x_r\ra$'s, but not a function of the phases of these amplitudes. For example, when $|x\ra$ is a single product state $|x_r\ra$, i.e., {\em Case A} in the main text, $\mathcal{B}=0$. Now the random phase is global, and does not lead to decoherence. If more than one coefficient is non-vanishing, so that $\mathcal{B} \neq 0$, there will be finite decoherence.

\begin{figure}[htbp]
\centering
\includegraphics[width=3.2in]{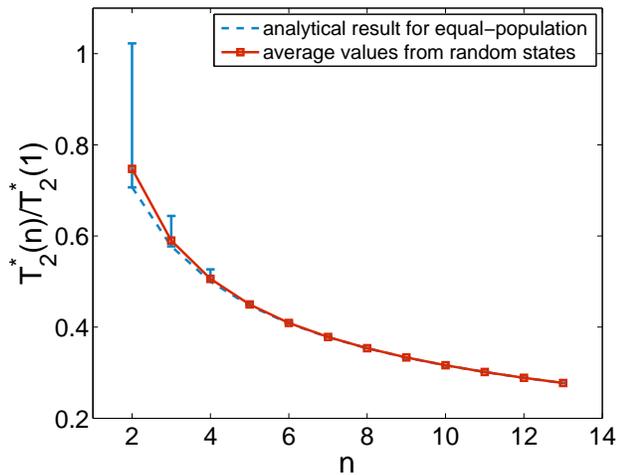}
\caption{The results of $T^*_2(n)/T^*_2(1)$ by randomly generated states over the whole Hilbert space of $n$-spin system, as compared with Eq.~(\ref{T2F}) in {\em Case F} in the main text. For each $n$, we use $100$ randomly populated states. }\label{errF}
\end{figure}

One can then use the expression of $B$ and Eq.~(\ref{AB}) to numerically obtain the scaling behavior of $T_2^*(n)/T_2^*(1)$ or $T_2(n)/T_2(1)$ for an arbitrary initial state. Although there is an infinite number of possible superposed states even for a small $n$, it turns out that the averages of the numerical results for different classes of states agree quite well with the analytical expressions obtained in the main text. For example, in Fig.~(\ref{errF}), we use $100$ randomly generated states with the bases of $|x\ra$ in {\em Case F} in the main text, i.e., all the product basis states in the whole Hilbert space, but random populations in each product basis state. The error bars in Fig.~\ref{errF} for random states rapidly vanishes with increasing $n$. Making the analytical expression a really good predictor of decoherence for an arbitrary state.

\end{document}